\documentstyle[12pt]{article}
\begin{document}

\title{Phase-coherence time saturation in mesoscopic systems: wave function
collapse }
\author{J. C. Flores}
\date{Universidad de Tarapac\'a, Departamento de F\'\i sica, Casilla 7-D, Arica,
Chile}
\maketitle

\bigskip

A finite phase-coherence time $\tau _\phi ^{meas}$ emerges from iterative
measurement onto a quantum system. For a rapid sequence, the phase-coherence
time is found explicitly. For the stationary charge conduction problem,  it
is bounded. At all order, in the time-interval of measurements, we propose a
general expression for $\tau _\phi ^{meas}$.

\bigskip

\bigskip


72.15.-v Electronic conduction in metals and alloys

03.65.Bz Theory of measurements

03.65.-w Quantum Mechanics

\newpage\ 

Since inelastic scattering is due to the absorption of quantum excitation
modes, is usually expected that phase breaking disappears a temperature $T=0$%
. In fact, it is well-known that electron-electron and electron-phonon
interaction produce a phase-coherence time behavior like $\tau _\phi \sim
1/T^p$ ($p>0$) (for a complete revision see [1,2]). So, at very low
temperature ($T\rightarrow 0$), the density operator $\widehat{\rho }$,
usually a mixture at non zero temperature, would be a pure state ($\widehat{%
\rho }=\widehat{\rho }^2$) with minimal lineal entropy $S=1-Tr\widehat{\rho }%
^2$. Surprisingly, in recent experiments in mesoscopic systems, it was
observed a complete saturation of $\tau _\phi $ at low temperature regime
[3-10]. This behavior seems quite general, observed in different
experimental realizations, and suggesting an intrinsic mechanism of
decoherence. Still when the explanation remains a controversial subject,
some theoretical papers explain this intriguing behavior by quantum
fluctuations of impurity ions [11], zero-point fluctuations of phase
coherent electrons [7,8], quantum fluctuations of the electric field in weak
localization theory [12]. A \ standard model explanation can be found in
[13]. For criticisms see [13-15].

$$
$$

In this paper, we propose an alternative (intrinsic) quantum mechanics
explanation to the observed saturation, namely, decoherence by \ iterative
measurement process (wave function collapse or reduction postulate).

$$
$$

In fact, ideal von Neumann's schema of measurement [16] onto a non
degenerate observable $\widehat{A},$ with discrete spectrum $a_l,$ 
\begin{equation}
\widehat{A}=\sum_{l=-\infty }^{l=+\infty }a_l\widehat{P_l},\quad \widehat{P_l%
}=|l><l|, 
\end{equation}
determines an irreversible change in the density operator $\widehat{\rho }$
given by [16-18] 
\begin{equation}
\widehat{\rho }^{\prime }=\sum_l\widehat{P_l}\widehat{\rho }\widehat{P_l}. 
\end{equation}
In this way, the non-diagonal terms $<l|\widehat{\rho }|s>$ ($l\neq s$) are
eliminated by the measurement and producing mixing ($\widehat{\rho }^{\prime
}\neq \widehat{\rho }^{\prime }{}^2$). We notice that $\widehat{\rho }%
^{\prime }$ is hermitian, positive, and satisfied the normalization
condition, provided that $\widehat{\rho }$ has these requirements.

$$
$$

As said before, to carrier-out explicit calculations, we consider an
iterative model of measurement. Namely, a set of $N$ ideal von Neumann's
measurements onto the observable $\widehat{A}$ (1) of the system. The
measurements are separated by a bounded interval of time $\Delta t$. Let $%
\widehat{\rho }$ be the density operator describing the state of the system
with Hamiltonian $\widehat{H}$. Using (2), the mapping between two
consecutive measurement is given by (see for instance [19,20]) 
\begin{equation}
\widehat{\rho }_{n+1}^{(+)}=\sum_l\widehat{P_l}e_{}^{i\widehat{H}\Delta
t/\hbar }\widehat{\rho }_n^{(+)}e^{-i\widehat{H}\Delta t/\hbar }\widehat{P_l}%
, 
\end{equation}
where $\widehat{\rho }_n^{(+)}$ stands for the density operator just after
the measurement at time $t_n=n\Delta t$. So, we have a free evolution ($%
n\Delta t^{+}\rightarrow (n+1)\Delta t^{-}$), carried out with the usual
unitary operator $\widehat{U}=exp(i\widehat{H}t/\hbar )$, and at the instant 
$t_{n+1}=(n+1)\Delta t$ operates the measurement process. 
$$
{} 
$$

From (3), after measurement, the density operator becomes diagonal in the
basis of $\widehat{A}$, i.e. 
\begin{equation}
\widehat{\rho }_n^{(+)}=\sum_lW_l^{(n)}\widehat{P_l}, 
\end{equation}
where $W_l^{(n)}$ is the probability to find the system in the state $l$ (at
time $t_n=n\Delta t)$. The evolution equation (3) can be written like to a
Markov-chain [19-20] for the probability $W$, explicitly,

\begin{equation}
W_l^{(n+1)}=\sum_{s=-\infty }^{s=+\infty }\left\| <l\mid e^{i\widehat{H}%
\Delta t/\hbar }\mid s>\right\| ^2W_s^{(n)}. 
\end{equation}
Nevertheless, we shall consider here only the case of small intervals of
time. \ Consider the well-known expansion 
\begin{equation}
e^{i\widehat{H}\Delta t/\hbar }\widehat{\rho }e^{-i\widehat{H}\Delta t/\hbar
}=\widehat{\rho }+i\frac{\Delta t}\hbar [\widehat{H},\widehat{\rho }]-\frac{%
\Delta t^2}{2!\hbar ^2}[\widehat{H},[\widehat{H},\widehat{\rho }]]+..., 
\end{equation}
which, at low order, gives the master equation for the probability $W$, 
\begin{equation}
W_l^{(n+1)}-W_l^{(n)}=\frac{\Delta t^2}{\hbar ^2}\sum_s||<l|\widehat{H}%
|s>||^2(W_s^{(n)}-W_l^{(n)}). 
\end{equation}
Namely, quadratic in the formal expansion parameter. The validity of the
expansion (6) will be discussed later. 
$$
{} 
$$

Assuming a hopping term only between nearest neighbors states, i.e.

\begin{equation}
<l\mid \widehat{H}\mid s>\sim b\delta _{s\pm 1,l}+b^{\prime }\delta _{s,l}, 
\end{equation}
the dispersion $\sigma _n^2$ , at time $n\Delta t$, in the $l$-space of
quantum number 
\begin{equation}
\sigma _n^2=\sum_ll^2W_l^{(n)}, 
\end{equation}
becomes related to the evolution equation : 
\begin{equation}
\sigma _{n+1}^2-\sigma _n^2=\frac{2b^2\Delta t^2}{\hbar ^2}. 
\end{equation}
Equation (10) allows to define the phase-coherence time $\ $in our model of
iterative measurement. In fact, (10) defines a diffusive motion in the $l$%
-space. If we start with a pure state $|l_o><l_o|$ with dispersion zero,
then ($t_n=n\Delta t$) 
\begin{equation}
\sigma _n^2=\frac{2b^2\Delta t}{\hbar ^2}t_n, 
\end{equation}
and mixture is produced after a finite time 
\begin{equation}
\tau _\phi ^{meas}=\frac{\hbar ^2}{2b^2\Delta t}. 
\end{equation}

The above relationship defines the phase-coherence time due to iterative
measurement onto the observable $\widehat{A}$ and it deserves some comments: 
$$
{} 
$$
(i) $\tau _\phi ^{meas}$ depends on the system properties (parameter $b$)
and the apparatus specifications (time $\Delta t$). However, wave function
collapse is an intrinsic propriety of quantum mechanics [16-18]. So, (12)
describes a quantum mechanics basic process. 
$$
{} 
$$
(ii) The formal expansion (6) requires that $b\Delta t/\hbar <<1$,
nevertheless decoherence exists at all order (see equation (17) below). 
$$
{} 
$$
(iii) The limit $\Delta t\rightarrow 0$ gives $\tau _\phi ^{meas}\rightarrow
\infty $ in accord with quantum Zeno effect [21]. In fact from (10),
diffusion in the $l$-space does not hold in this limit due to the
measurement process. 
$$
{} 
$$
(iv) For dissipative open systems (charge conduction, absorption, etc.) in
the stationary regime, the relaxation time $\tau _\gamma $ must be smaller
than $\Delta t,$ i.e. 
\begin{equation}
\tau _\gamma <\Delta t.
\end{equation}
If (13) is not verified, dissipation to environment is not directly possible
because we approach the regime (iii). More important, from equation (13) we
have a bound for the decoherence time (12) given by 
\begin{equation}
\tau _\phi ^{meas}<\frac{\hbar ^2}{2b^2\tau _\gamma },
\end{equation}
and only valid \ in the stationary regime. 
$$
{} 
$$
(v) Random independent measurements intervals $\Delta t_n$ , with small
fluctuations [19], do not change the definition (12). So, a similar
definition for the coherence time \ would be made here. 
$$
{} 
$$

The case of charge conduction (iv) can be put in other way. Consider the
Drude relationship between the relaxation time $\tau _\gamma $ and the
classical conductivity $\sigma :$ 
\begin{equation}
\tau _\gamma =\frac{m\sigma }{\delta nq^2}, 
\end{equation}
where $\delta n$ is the number of carriers of mass $m$ and charge $q$. Then,
(14) can be written as 
\begin{equation}
\tau _\phi ^{meas}<\frac{\hbar ^2q^2\delta n}{2b^2m\sigma }, 
\end{equation}
namely, in the stationary conduction problem with dissipation, the coherence
time due to the measurement process is bounded. The above expression can be
re-write in function of the diffusion constant $D$ by using the Einstein
relation $\sigma =Dq^2(dn/dE)$ [2,8] and we obtain $\tau _\phi ^{meas}<\frac{%
\hbar ^2}{2mD}\frac{\delta E}{b^2}$,where $\delta E$ denotes the energy-wide
of the carriers.

\bigskip

Expression (12) for the coherence time was defined using the expansion (6).
Namely, it is valid in the limit of small intervals of time (ii) and small
hopping (8). Nevertheless, \ it can be \ extended at all order from the
definition of the probability $W$. In fact, we ask about the time necessary
to lose coherence, when the initial \ state is \ a pure-state (for instance $%
|l_0><l_0|$). \ So, this suggests the general definition at all order in \ $%
\Delta t$:

\begin{equation}
\frac 1{\tau _\phi ^{meas}}=-\lim _{n\rightarrow \infty }\frac 1{2n\Delta
t}\ln \left( W_{l_0}^{(n)}\right) , 
\end{equation}
which must be independent of the state $l_0.$ The first order \ calculation
on (17) coincides with (12) and showing the independence of the initial
state $l_0$. $W_{l_0}^{(n)}$ in (17), can be related to $W_{l_0}^{(0)}=%
\delta _{l,l_0}$ by the recursion rule (5) (Markov-chain). In fact, $1/\tau
_\phi ^{meas}$ corresponds to the so-called (minimum) Lyapounov exponent of
the system.

$$
$$

In conclusion, we have considered an iterative model of measurement (3,5)
onto an observable of a given quantum system. In the limit of small
intervals of time $\Delta t$ (ii) and hopping between nearest neighbor (8),
the master equation associated to the evolution probability (7) allows to
define the phase-coherence time $\tau _\phi ^{meas}$ (12). So, necessarily,
iterative measurement on a quantum system produce intrinsic decoherence. In
the case with dissipation (stationary regime), the phase-coherence time is
bounded (14,16) and no divergent. For all order in $\Delta t$, we suggest
the definition (17) for $\tau _\phi ^{meas}$.

\bigskip

Acknowledges: Useful discussions related to the phase-coherence time were
carried out with Professor V. Bellani (Pavia University) and, respect to
wave function collapse, with Professor S.\ Montecinos (UFRO). This work was
partially realized in Pavia University (CICOPS fellowship) and partially
supported by FONDECYT.

\end{document}